\documentstyle[aps,twocolumn]{revtex}
\newcommand{\deltabar}{\,\,{\bar{}\hspace{1pt}\! \!\delta }}

\begin{document}
\title{Autoparallels From a New Action Principle}
\author{H. Kleinert and A. Pelster}
\address{Institut f\"ur Theoretische Physik,
Freie Universit\"at Berlin, Arnimallee 14, 14195 Berlin, Germany}
\date{\today}
\maketitle
\begin{abstract}
We present a simpler and more powerful version of the
recently-discovered action principle for the
motion of a spinless point particle in spacetimes
with curvature and torsion.
The surprising feature of the new
principle is that an
action
involving only the metric can produce an equation of motion
with a torsion force, thus changing geodesics to
autoparallels. This additional torsion
force arises from a noncommutativity
of variations with
parameter derivatives of the paths due to
the closure failure of parallelograms in the presence of torsion.\\

\end{abstract}
According to Einstein's equivalence principle,
the equation of motion of a free spinless
point particle
in a curved spacetime
is found by the following two step procedure.
First, the equation of motion
\begin{equation}
\label{NEWTON}
 \ddot{x}^{\,a} ( \tau ) \, = \, 0
\end{equation}
for rectilinear coordinates
$x^{\,a}$ $(a = 0 , 1 , 2 , 3 )$
in a flat spacetime with
the Minkowski metric
$( \eta_{ab} ) = ( + , - ,
- , - )$ is transformed via
\begin{equation}
\label{GLOBAL}
x^{\,a} \, =\, x^{\,a} ( q )
\end{equation}
to curvilinear coordinates $q^{\,\lambda}$
$(\lambda = 0 , 1 , 2 , 3)$.
The resulting equation reads
\begin{equation}
\label{GEO}
\ddot{q}^{\,\lambda} ( \tau ) \, +
\overline{\Gamma}_{\mu\nu}^{\,\,\,\,\,\lambda} ( q ( \tau ) ) \,
\dot{q}^{\,\mu} ( \tau ) \, \dot{q}^{\,\nu} ( \tau ) \,
= \, 0 \, ,
\end{equation}
where $\tau $ is the proper time, and  the Christoffel connection
\begin{eqnarray}
&&\overline{\Gamma}_{\mu\nu}^{\,\,\,\,\,\lambda} ( q ) \, = \,
\frac{1}{2} \, g^{\lambda \kappa} ( q ) \,
\nonumber\\
&&\label{CHRISTOFFEL}
\times  \, \Big[ \partial_{\mu} g_{\nu\kappa} ( q ) \,
+ \, \partial_{\nu}
g_{\kappa\mu} ( q ) \, - \,
\partial_{\kappa} g_{\mu\nu} ( q ) \Big]
\end{eqnarray}
is derived from the induced metric
\begin{eqnarray}
g_{\mu\nu} ( q ) & = & e^{\,a}_{\,\,\,\mu} ( q )
e^{\,b}_{\,\,\,\nu} ( q ) \eta_{ab} \, , \label{IN}\\
e^{\,a}_{\,\,\,\lambda} ( q ) & \equiv & \partial x^a ( q )
 / \partial q^{\lambda}. \label{DER}
\end{eqnarray}
Second, it
is postulated that (\ref{GEO}) and (\ref{CHRISTOFFEL}) also
describe the motion
in a curved spacetime with an intrinsic metric $g_{\mu \nu}$.
The solutions of (\ref{GEO}) represent the shortest curves
in the spacetime, i.e. {\em geodesics\/}. \\

If torsion
is admitted to the geometry, thus generalizing  the
Riemann to a Cartan spacetime \cite{Schouten},
there exists
another equation of motion,
which is as covariant and simple
as (\ref{GEO}),
\begin{equation}
\label{AUTO}
\ddot{q}^{\,\lambda} ( \tau ) \, +
\Gamma_{\mu\nu}^{\,\,\,\,\,\lambda} ( q ( \tau ) ) \,
\dot{q}^{\,\mu} ( \tau ) \, \dot{q}^{\,\nu} ( \tau ) \, = \, 0 \, ,
\end{equation}
in which
the
Christoffel connection
$\overline{\Gamma}_{\mu\nu}^{\,\,\,\,\,\lambda}$ is
replaced by the full
affine or Cartan connection
\begin{equation}
\label{AFFINE}
\Gamma_{\mu\nu}^{\,\,\,\,\,\lambda} ( q ) \, = \,
\bar \Gamma_{\mu\nu}^{\,\,\,\,\,\lambda} ( q ) \, + \,
K_{\mu\nu}^{\,\,\,\,\,\lambda} ( q ) \, .
\end{equation}
It contains in addition to the Christoffel connection (\ref{CHRISTOFFEL})
the contortion tensor
\begin{equation}
\label{CONT}
K_{\mu\nu}^{\,\,\,\,\,\lambda} ( q ) \, = \,
S_{\mu\nu}^{\,\,\,\,\,\lambda} ( q ) \, - \, S_{\nu\,\,\,\mu}^{\,\,\lambda}
( q ) \, +\, S^{\lambda}_{\,\,\,\mu\nu} ( q ) \, ,
\end{equation}
a combination of torsion tensors
\begin{equation}
\label{TORDEF}
S_{\mu\nu}^{\,\,\,\,\,\lambda} ( q )
\equiv  \frac{1}{2} \Big[ \Gamma_{\mu\nu}^{\,\,\,\,\,\lambda} ( q ) -
\Gamma_{\nu\mu}^{\,\,\,\,\,\lambda} ( q )  \Big] \, .
\end{equation}
The solutions of (\ref{AUTO}) define
the straightest
curves in the spacetime, i.e. {\em autoparallels}.
Note that due to (\ref{AFFINE}) and (\ref{CONT})
the autoparallels (\ref{AUTO}) and the geodesics
(\ref{GEO}) differ by the additional torsion force
$K_{\mu\nu}^{\,\,\,\,\,\lambda}(q)\dot{q}^{\mu}\dot{q}^{\nu} =
2S^ \lambda{}_{\mu \nu}(q)\dot{q}^{\mu}\dot{q}^{\nu}$.
It should be noted, that the symmetric part
$K_{(\mu\nu)}^{\,\,\,\,\,\,\,\,\,\,\lambda}$
of the contorsion tensor does {\em not} identically vanish.\\

The question arises which of the two curves provides us with the
correct particle trajectories in spacetimes with curvature and torsion.
According to the Einstein-Cartan theory of gravity
\cite{Hehl1,Kleinert3,Hehl2},
the canonical energy-momentum tensor $\Theta ^{\mu \nu}$
and the spin density $ \Sigma^{\mu \nu, \lambda}$
of matter
should determine the geometry
of spacetime by equations $G^{\mu \nu}= -\kappa\Theta ^{\mu \nu}$
and   $ S^{\mu \nu, \lambda}=- \kappa \Sigma^{\mu \nu, \lambda}$,
where $ \kappa$ is the gravitational constant,
 $S^{\mu \nu, \lambda}$ the Palatini tensor derived from the
torsion tensor $S_{\mu\nu}^{\,\,\,\,\,\lambda}$
and   $G^{\mu \nu}$
the Einstein tensor formed from the Cartan curvature tensor
\begin{eqnarray}
&&\!R_{\mu\nu\kappa}^{\,\,\,\,\,\,\,\,\,\,\lambda} ( q )  =   \partial_{\mu}
\Gamma_{\nu\kappa}^{\,\,\,\,\,\,\lambda} ( q ) - \partial_{\nu}
\Gamma_{\mu\kappa}^{\,\,\,\,\,\,\lambda} ( q ) \nonumber \\
&&~~~~~~~ -
\Gamma_{\mu\kappa}^{\,\,\,\,\,\,\rho} ( q )
\Gamma_{\nu\rho}^{\,\,\,\,\,\,\lambda} ( q )
+ \Gamma_{\nu\kappa}^{\,\,\,\,\,\,\rho} ( q )
\Gamma_{\mu\rho}^{\,\,\,\,\,\,\lambda} ( q )\, . \label{RR}
\end{eqnarray}
The Bianchi identity
for the Einstein tensor $G^{\mu \nu}$
 implies for the symmetric energy-momentum tensor $T ^{\mu \nu}$
of spinless point particles the conservation law
\begin{equation}
\bar D_{ \nu}T^{\mu \nu} ( q ) = 0 \, ,
\label{Tcons@}\end{equation}
where $\bar D_{ \nu}$ is the covariant derivative
involving
the Christoffel connection $\overline{\Gamma}_{\mu\nu}^{\,\,\,\,\,\lambda}$.
This conservation law does not contain the torsion tensor
$S_{\mu\nu}^{\,\,\,\,\,\lambda}$,
and leads therefore directly to the
geodesic equation  (\ref{GEO}) as shown by Hehl \cite{Hehl3}.\\

Because of the beauty
of the mathematical framework within which this result was derived,
there was little doubt that particles should run along the shortest paths,
just as in Einstein's original theory of gravity.
Physically, however, it is hard to conceive how this can be true,
since it contradicts
two quite fundamental properties of physical laws: inertia and locality.
Because of its inertia, a particle will change its
direction in a minimal way at each instant of time,
which makes its trajectory as straight as possible.
If it were to choose a path which
minimizes the length of the orbit it would have
possessed some global information of the geometry.
In Einstein's theory of gravity, the two paths
happen to coincide on mathematical grounds, so that
this basic problem
did not become apparent for geodesics, but in the presence of torsion
it can no longer be ignored.\\

Doubts as to the correctness of geodesics
as particle trajectories arose first in a completely
different context \cite{Kleinert1,Kleinert2}.
When solving nonrelativistic Coulomb
problems in classical or quantum
mechanics,
a local coordinate transformation
\begin{equation}
\label{LCT}
d x^{\,a} \, = \, e^{\,a}_{\,\,\,\lambda} ( q ) \, d q^{\,\lambda}
\end{equation}
with the Kustaanheimo-Stiefel coefficients \cite{Kust1,Kust2}
\begin{equation}
\left( e^{\,a}_{\,\,\,\lambda} ( q ) \right) =
\left( \begin{array}{rrrr}
q^3 & q^0 & q^1 & q^2 \\
q^0 & - q^3 & - q^2 & q^1 \\
q^1 & q^2 & - q^3 & - q^0 \\
q^2 & - q^1 & q^0 & - q^3
\end{array} \right)
\label{COEFF}
\end{equation}
has long been  very helpful since it carries  Coulomb into
 harmonic systems.
The
coefficient functions
$e^{\,a}_{\,\,\,\lambda}$
possess the interesting
property of
not obeying  the integrability
condition of Schwarz:
\begin{eqnarray}
\partial_{\mu} \, e^{\,a}_{\,\,\,\lambda} ( q ) \, - \,
\partial_{\lambda}\, e^{\,a}_{\,\,\,\mu} ( q ) & \neq & 0 \, .
\label{SCHWARZ1}
\end{eqnarray}
This implies that
there exists no
singlevalued global transformation (\ref{GLOBAL})
from which
$e^{\,a}_{\,\,\,\lambda}$ could be obtained as
the derivatives (\ref{DER}).
This makes the local coordinate transformation (\ref{LCT}) nonholonomic.
In the absence of forces in the original flat space, particles
run along straight lines which satisfy
the equation (\ref{NEWTON}).
Their image under the transformation (\ref{LCT})
satisfies the autoparallel equation (\ref{AUTO}),
where the affine connection is explicitly given by
\begin{equation}
 \Gamma_{\mu\nu}^{\,\,\,\,\,\lambda} ( q )
=
 e_{a}^{\,\,\,\lambda} ( q )
 \partial_\mu  e^{a}{}_{\nu} ( q ) .
\label{Gamma1}
\end{equation}
{}From (\ref{TORDEF}), (\ref{RR}) and (\ref{Gamma1})
we then conclude that
a coordinate transformation
of the type
(\ref{LCT})
carries a flat space into a space with the torsion tensor
\begin{equation}
\label{TOREX}
S_{\mu\nu}^{\,\,\,\,\,\lambda} ( q )
= \frac{1}{2} e^{a}_{\,\,\,\lambda} ( q )
 \Big[
\partial_{\mu} e^{a}_{\,\,\,\nu} ( q ) -
\partial_{\nu}  e^{a}{}_{\mu} ( q ) \Big]
\end{equation}
and a vanishing Cartan
curvature tensor \cite{Kleinert2}.
Now, it is well-known that
equations of motion remain valid
under such mappings. Therefore
autoparallels must be the correct
particle trajectories \cite{Kleinert1,Kleinert2}.\\

It is not hard to generate also
nonvanishing Cartan curvature
by nonholonomic mappings
(\ref{LCT}).
For this, the coefficient functions
$e^{a}{}_{\lambda}$
must only be chosen as multivalued.
Then the functions $e^{a}{}_{\lambda}$ themselves
fail to satisfy the criterion of Schwarz,
and the noncommutativity of partial derivatives
yields
for the Cartan curvature tensor
(\ref{RR}) the expression
\begin{eqnarray}
R_{\mu \nu \lambda}{}^ \kappa ( q ) =
e_{\,a}^{\,\,\, \kappa} ( q )
\Big[ \partial_{\mu} \partial_{\nu}-
\partial_{\nu} \partial_{\mu}
\Big] \, e^{\,a}_{\,\,\,\lambda} ( q ) \, ,
\label{SCHWARZ}
\end{eqnarray}
as can be verified by inserting (\ref{Gamma1}).
Local coordinate transformations (\ref{LCT}) which generate
both torsion (\ref{TOREX}) and curvature (\ref{SCHWARZ})
are widely used for describing crystals with defects
\cite{Kondo,Bilby,Kroener1,Kroener2}.
Examples illustrating such mappings and their applications
are elaborated in \cite{Kleinert3}.\\

In Ref.~\cite{Fiziev1}
it was pointed out that autoparallel trajectories
could only be understood
after a revision of the variational
calculus in spacetimes with torsion.
These possess an unusual feature not encountered
before, namely
a {\em closure failure of parallelograms\/}.
As a consequence,
variations of particle trajectories
in an action cannot be performed as usual.
Variations $ \delta x^a(\tau)$ of a path in flat
spacetime are always
performed at vanishing endpoints, thus forming closed paths.
The images $ \delta q^\lambda(\tau)$
under a nonholonomic mapping
(\ref{LCT}), however, are in general open, their
closure failure being proportional to
the torsion. The superscript $S$ indicates this special feature.\\

Let us
briefly recall
the derivation in   \cite{Fiziev1}.
If $e_{\,a}^{\,\,\,\lambda}$ denotes
the reciprocal multivalued basis vectors
orthonormal to
$e^{\,a}_{\,\,\,\mu}$,
\begin{equation}
e_{\,a}^{\,\,\,\lambda} ( q ) \, e^{\,a}_{\,\,\,\mu} ( q ) \, = \,
\delta^{\,\lambda}_{\,\,\,\mu} \, ,
\end{equation}
the local coordinate transformation (\ref{LCT}) can be integrated
implicitly for any orbit  $x^{\,a}(\tau)$ in the flat spacetime
to yield an implicit equation for the image orbit $q^{\lambda} ( \tau )$
in the spacetime with curvature and torsion:
\begin{equation}
\label{INTEGRAL}
\hspace*{-0.2cm}
q^{\,\lambda} ( \tau )  = q^{\,\lambda} ( \tau_0 ) +
\int\limits_{\tau_0}^{\tau}d \tau' e_{\,a}^{\,\,\,\lambda}
\left( q ( \tau' ) \right) \dot{x}^{\,a} ( \tau' ) .
\end{equation}
If $\delta x^{\,a}(\tau)$ denotes an arbitrary variation of
the orbits in the flat spacetime,
nonholonomic variations $\delta^S q^{\,\lambda}
(\tau)$ are defined by identifying
the image
of the varied path $x^{\,a} (\tau)+ \delta x^{\,a} (\tau)$
under the mapping
(\ref{INTEGRAL}) with
$q^{\,\lambda} (\tau)+\delta^S q^{\,\lambda} (\tau)$.
The variations
$\delta x^{\,a} ( \tau )$ are performed as usual at fixed
end points. Their images $ \delta^S q^\lambda ( \tau )$, however,
possess
the above-mentioned
closure failure which is proportional
to the amount of torsion introduced inside $ \delta^S q^\lambda ( \tau )$ by
the nonholonomic coordinate transformation (\ref{LCT}).
They are {\em open nonholonomic variations\/} which
may be chosen to vanish at the initial point, but then
they are nonzero
at the final point.
Note that the
open nonholonomic variations $\delta^S$ commute with the parameter derivatives
$d_\tau \equiv \partial
/\partial \tau$:
\begin{equation}
\label{NC}
\delta^S d_\tau q^{\,\lambda} -
d_\tau \delta^S q^{\,\lambda}  =  0  .
\end{equation}
Applying these variations to an action
\begin{equation}
{\cal A} [ q^{\,\lambda} ( \tau ) ]=
\int\limits_{\tau_1}^{\tau_2}d\tau L(q(\tau),\dot
q(\tau)),
\label{AC@}\end{equation}
the correct variational principle
in the presence of curvature and torsion
was found in \cite{Fiziev1} to have the form
\begin{equation}
\label{FI}
\delta^S {\cal A} [ q^{\,\lambda} ( \tau ) ] \, = \, 0 \, .
\end{equation}
Applying
this new variational principle to the Lagrangian of
a spinless point particle
\begin{equation}
\label{FREE}
L ( q , \dot{q} ) = - M c \sqrt{g_{\lambda \mu} ( q ) \dot{q}^{\lambda}
\dot{q}^{\mu}}
\end{equation}
produces directly
the autoparallel equation  (\ref{AUTO}).\\

In spite of the simplicity of the result,
the
algebra involved in deriving the torsion terms
in the equation of motion (\ref{AUTO})
turned out to be
quite complicated \cite{Fiziev1}.
In addition, the applicability of the
procedure
was restricted to a free particle  Lagrangian,
and did not permit the inclusion of nongeometric forces
such as electromagnetic ones.
The purpose
of this note is
to improve this situation by presenting a
variational procedure
which is more elegant
and can be applied to general Lagrangians.
This is possible by avoiding
the awkward open anholonomic variations
$\delta^Sq^\mu(\tau)$ in favor of {\em auxiliary closed\/}
nonholonomic variations $\deltabar$ which do
vanish at the endpoints, in this respect being
closer to the ordinary variations.
They are {\em defined\/} as the images of
the ordinary variations $\delta x^{\,a}(\tau)$ in the flat
spacetime
under the local coordinate transformation (\ref{LCT}):
\begin{equation}
\label{HV}
\deltabar q^{\,\mu} (\tau)\, \equiv \, e_{\,a}^{\,\,\,\mu} ( q(\tau) ) \,
\delta x^{\,a} (\tau)\, .
\end{equation}
Their special property, which
will generate the torsion force later on,
is that
they do not
commute with parameter
derivatives of the path functions $q^ \mu(\tau)$.
To see
this we invert (\ref{HV}) and take the parameter derivative
to obtain
\begin{equation}
{d_\tau} \, \delta x^{\,a} = e^{\,a}_{\,\,\,\lambda} ( q )
d_\tau \deltabar q^{\,\lambda}
\label{COM1}
+ \partial_{\mu} e^{\,a}_{\,\,\,\lambda} ( q ) \dot{q}^{\,\mu}
 \deltabar q^{\,\lambda} .
\end{equation}
A variation of
(\ref{LCT}) yields, on the other hand,
\begin{equation}
\delta d_\tau x^{\,a} \, = \, e^{\,a}_{\,\,\,\lambda}
( q ) \deltabar
\dot q^{\,\lambda}
\label{COM2}
+  \partial_{\mu}
 e^{\,a}_{\,\,\,\lambda} ( q ) \, \dot{q}^{\,\lambda} \, \deltabar
q^{\,\mu} \, .
\end{equation}
Using now the property
\begin{equation}
\delta  {d_\tau } x^{\,a}
-  {d_\tau } \delta x^{\,a}
 =  0
\end{equation}
in  flat spacetime,
we deduce from (\ref{COM1}) and (\ref{COM2}) that
the operations $ \deltabar$ and $d_\tau$
satisfy the commutation relation
\begin{equation}
\label{COMMUTE}
\deltabar d_\tau q^{\lambda}
- d_\tau \deltabar q^{\lambda}
  =
 2 \, S_{\mu\nu}^{\,\,\,\,\,\lambda} ( q ) \,
\dot{q}^{\mu} \, \deltabar q^{\,\nu} \, .
\end{equation}
The new action principle
states now
\begin{equation}
\label{HP1}
\deltabar {\cal A}
[ q^{\,\lambda} ( \tau ) ] \, = \, 0
\end{equation}
for all variations $\deltabar q^{\,\lambda} ( \tau )$
which vanish at the end points $\tau_1$ and $\tau_2$,
\begin{equation}
\label{HP2}
\deltabar q^{\,\lambda} ( \tau_1 ) \, = \, \deltabar q^{\,\lambda}
( \tau_2 ) \, = \, 0 \, ,
\end{equation}
and which satisfy the commutation relation
(\ref{COMMUTE}).
The latter property of the auxiliary
variations in spacetimes with torsion
modifies the Euler-Lagrange equations by the correct torsion force.
Indeed, varying
the action
(\ref{AC@}) by $ \deltabar q^\lambda (\tau)$
yields
\begin{eqnarray}
&&
\!
\!\!\!\!\!\!
\deltabar {\cal A} [ q^{\,\lambda} ( \tau ) ]=\nonumber \\
 & =&  \int\limits_{\tau_1}^{\tau_2} d \tau
 \left[ \frac{\partial L}{\partial q^{\lambda} ( \tau )}
 \deltabar  q^{\lambda} ( \tau )
+ \frac{\partial L}{\partial \dot{q}^{\lambda}
( \tau )}  \frac{d}{d \tau}  \deltabar q^{\lambda} ( \tau ) \right.
 \nonumber \\
\label{VARIATION}
&& \left. +  2 \,S_{\mu\nu}^{\,\,\,\,\,\lambda} \left( q ( \tau )
\right)
 \frac{\partial L}{\partial \dot{q}^{\lambda} ( \tau )} \,
\dot{q}^{\mu} ( \tau )  \deltabar q^{\nu} ( \tau ) \right]   .
\end{eqnarray}
After a partial integration of the second term,
we obtain
the modified
Euler-Lagrange equation
\begin{eqnarray}
\frac{\partial L}{\partial q^{\,\lambda} ( \tau )} \,
- \,
\frac{d}{d t} \, \frac{\partial L}{\partial
\dot{q}^{\,\lambda} ( \tau )}
\nonumber \\
\label{EL}
= 2  S_{\lambda\mu}^{\,\,\,\,\,\nu} \left( q ( \tau ) \right)
\dot{q}^{\mu} ( \tau ) \frac{\partial
L}{\partial \dot{q}^{\nu} ( \tau )}
\, ,
\end{eqnarray}
the right-hand side being the torsion force.\\

For the free-particle Lagrangian (\ref{FREE}),
the equation of motion (\ref{EL}) takes the form
\begin{eqnarray}
\hspace*{-1cm}
\ddot{q}^{\lambda} \Big[ g^{\lambda \kappa}  \frac{1}{2} \Big( \partial_{\mu}
g_{\nu\kappa}   + \partial_{\nu} g_{\kappa\mu}
\nonumber \\
-  \partial_{\kappa}
g_{\mu\nu}   \Big)
+  2 S^{\,\lambda}_{\,\,\,\mu\nu}\Big]\,\dot{q}^{\,\mu} \, \dot{q}^{\nu} =
0,
\end{eqnarray}
which coincides with the
autoparallel equation (\ref{AUTO}) after
taking into account (\ref{CHRISTOFFEL}), (\ref{AFFINE}) and (\ref{CONT}).
Thus spinless point particles
move along straightest lines,
as expected from particle inertia and locality,
rather than shortest lines, as commonly believed. Furthermore
we observe the increased power of the new variational principle
(\ref{HP1}).
The modified Euler-Lagrange equation
(\ref{EL}) make also sense
if the Lagrangian
contains
a coupling to nongeometric fields, as for instance the vector potential
$A_{\mu}$.\\

The new action principle
 presented here
is the most natural generalization of the fundamental Hamilton
principle
to spacetimes with curvature and torsion.
In contrast to \cite{Fiziev1},
variations can be performed without reference
to the
flat
coordinate system, which was only introduced as a mathematical
device
for
{\em deriving\/}
the crucial commutation relation (\ref{COMMUTE})
between
variation and parameter derivative. \\

Since spinless point particles
run along autoparallel trajectories
on very fundamental grounds , i.e. inertia, locality and closure failure,
the gravitational field equations in spaces with torsion, on which
the presently accepted derivation of geodesic trajectories
is based, must have a fundamental flaw.
In particlular, they must be such that the covariant conservation law
for the energy-momentum tensor of the point particle is
\begin{equation}
\bar D_ \nu T^{\mu \nu}+2 S^\mu{}_{ \lambda \kappa}T^{ \kappa
\lambda}=0 \, ,
\end{equation}
instead of (\ref{Tcons@}).
This will be an important task for the future.\\

{\bf Note added in proof:}\\

Between the writing of this letter and its
final print in
Gen. Rel. Grav. {\bf 31}, 1439 (1999), numerous new results have been found.
For gradient torsion, gravitational field equations have been set up
[1,2,3] whose semiclassical trajectories are
autoparallels [4]. However, it seems to be impossible to construct a
realistic field theory with a general torsion field. First, there are
serious consistency problems of the
coupling of massless versus massive vector mesons to torsion [4].
Second, field theories with general torsion seem to violate a basic
universality principle of angular and spin momentum in elementary
particle physics [5].\\

\begin{tabular}{@{}lp{6.6cm}}
 $[1]$ & H. Kleinert, {\em Nonholonomic Mapping Principle for Classical
and Quantum Mechanics in Spaces with Curvature and Torsion}
(aps1997sep03 002).\\
 $[2]$ & H. Kleinert, Act. Phys. Pol. {\bf B 29}, 1033 (1998)
(gr-qc/9801003).\\
 $[3]$ & H. Kleinert, A. Pelster, Act. Phys. Pol. {\bf B 29},
1015 (1998) (gr-qc/9801030). \\
 $[4]$ & H. Kleinert, Phys. Let {\bf B 440}, 283 (1998)
(gr-qc/9808022).\\
 $[5]$ & H. Kleinert, {\em Universality Principle for Orbital Angular}
(gr-qc/9807021).
\end{tabular}

\end{document}